\documentclass[aps,pra,preprint,floatfix,superscriptaddress]{revtex4-1}
\usepackage{amsmath, amssymb, graphics, mathrsfs, bm, stackrel,bbold}
\usepackage[oldstylenums]{kpfonts}
\usepackage[cal=boondoxo]{mathalfa}
\usepackage[usenames,dvipsnames]{xcolor}
\definecolor{Highlight}{rgb}{1,1,0.75}
\newcommand{\authormain}{Srividya Iyer-Biswas}
\newcommand{\titlemain}{Bridging the timescales of single-cell and population dynamics}
\usepackage{graphicx}
\usepackage{epstopdf}
\usepackage[bookmarks={false}, pdfauthor={\authormain}, pdftitle={\titlemain}]{hyperref}
\hypersetup{colorlinks=true, linkcolor=BrickRed, citecolor=Violet, filecolor=OliveGreen, urlcolor=RoyalBlue, filebordercolor={.8 .8 1}, urlbordercolor={.8 .8 0}}
\usepackage[all]{hypcap}

\usepackage[export]{adjustbox}
\newcommand\ba{\begin{array}}
\newcommand\ea{\end{array}}
\newcommand\nn{\nonumber}

\newcommand\ri{\right}
\renewcommand\le{\left}

\renewcommand\a{\alpha}

\renewcommand\d{\delta}

\newcommand\f{\phi}

\newcommand\g{\gamma}

\renewcommand\k{\kappa}

\newcommand\m{\mu}

\newcommand\n{\nu}

\newcommand\rr{\rho}

\newcommand\s{\sigma}

\renewcommand\t{\tau}

\newcommand\imply{\Rightarrow}

\newcommand\la{\langle}
\newcommand\ra{\rangle}
\newcommand\pd{\partial}

\begin{document}
\title{\titlemain}
\author{Farshid Jafarpour}
\affiliation{Department of Physics and Astronomy, Purdue University, West Lafayette, IN 47907}
\author{Charles S. Wright}
\affiliation{Department of Physics and Astronomy, Purdue University, West Lafayette, IN 47907}
\author{Herman Gudjonson}
\affiliation{James Franck Institute and Institute for Biophysical Dynamics, University of Chicago, Chicago, IL 60637}
\author{Jedidiah Riebling}
\affiliation{Department of Physics and Astronomy, Purdue University, West Lafayette, IN 47907}
\author{Emma Dawson}
\affiliation{Department of Physics and Astronomy, Purdue University, West Lafayette, IN 47907}
\affiliation{Department of Physics, St Olaf College, Northfield, MN, 55057}
\author{Klevin Lo}
\affiliation{Department of Biochemistry and Molecular Biology, University of Chicago, Chicago, IL 60637}
\author{Aretha Fiebig}
\affiliation{Department of Biochemistry and Molecular Biology, University of Chicago, Chicago, IL 60637}
\author{Sean Crosson}
\affiliation{Department of Biochemistry and Molecular Biology, University of Chicago, Chicago, IL 60637}
\author{Aaron R. Dinner}
\affiliation{James Franck Institute and Institute for Biophysical Dynamics, University of Chicago, Chicago, IL 60637}
\author{\authormain}
\email{iyerbiswas@purdue.edu}
\affiliation{Department of Physics and Astronomy, Purdue University, West Lafayette, IN 47907}
\affiliation{Santa Fe Institute, Santa Fe, NM 87501}

\begin{abstract}
How are granular details of stochastic growth and division of individual cells reflected in smooth deterministic growth of population numbers? We provide an integrated, multiscale perspective of microbial growth dynamics by formulating a data-validated theoretical framework that accounts for observables at both single-cell and population scales. We derive exact analytical complete time-dependent solutions to cell-age distributions and population growth rates as functionals of the underlying interdivision time distributions, for symmetric and asymmetric cell division. These results provide insights into the surprising implications of stochastic single-cell dynamics for population growth. Using our results for asymmetric division, we deduce the time to transition from the reproductively quiescent (swarmer) to replication-competent (stalked) stage of the {\em Caulobacter crescentus} lifecycle. Remarkably, population numbers can spontaneously oscillate with time. We elucidate the physics leading to these population oscillations. For {\em C. crescentus} cells, we show that a simple measurement of the population growth rate, for a given growth condition, is sufficient to characterize the condition-specific cellular unit of time, and thus yields the mean (single-cell) growth and division timescales, fluctuations in cell division times, the cell age distribution, and the quiescence timescale.
\end{abstract}
\maketitle

\section{Introduction}

Several decades ago, the earliest quantitative microbiology experiments revealed that microbial population sizes increase exponentially under favorable growth conditions~\cite{monod}. However, direct quantification of the underlying growth and division dynamics at the level of the individual cell has only recently become possible, following advances in quantitative single-cell technologies~\cite{2014-iyer-biswas-mz, johan, kussell, gore, josh, naama, dan, strovas, wakamoto}. While population growth of cells is typically a deterministic process, which follows a smooth exponential function under favorable conditions, single-cell growth and division dynamics are highly stochastic. For instance, there is significant stochasticity in the division times (also known as generation times, cell lifetimes, interdivision times, or waiting times)---typically the COV (the coefficient of variation, defined as the ratio of the standard deviation to the mean) of division times is $10-30\%$~\cite{2014-iyer-biswas-mz, johan, kussell, gore, josh, naama, dan, strovas, talia, wakamoto, powell, kelly, weibel}. Using recent advances in single-cell technologies, it is possible to characterize these fluctuations with exquisite precision for large ensembles of statistically identical cells~\cite{2014-iyer-biswas-mz}.

In this paper we take advantage of the availability of high-quality quantitative single-cell datasets and develop an integrated perspective of microbial growth dynamics under balanced conditions. We formulate a data-driven theoretical framework that takes into account observables at both single-cell and population scales, and derive exact analytical complete time-dependent solutions to cell-age distributions and population growth rates as functionals of the underlying interdivision time distributions, for symmetric and asymmetric cell division. These results provide insights into the surprising implications of stochastic single-cell dynamics for population growth.

Exponential growth dynamics are also observed (to a good approximation) in the population growth of other species, the spreading of pandemics, the polymerase chain reaction for amplification of DNA fragments, global internet traffic increase, multiplication of viruses in T cells, and tumor growth~\cite{2010-hagen-pb}. There are unifying themes in the quantitative characterizations of exponential growth in these different contexts, codified in general results of the theory of branching processes~\cite{2008-feller-hc, 2002-kimmel-th}. The idiosyncrasies of each experimental system determine the important dynamical variables and measurables in that specific context.  For instance, in demographic studies the age structure of the human population in a region at a given time can be well characterized. However, typically there is a paucity of data for the total population number over multiple human lifespans. Thus indirect estimation of the Malthusian parameter (the exponential growth rate) from the observed age structures is a focus of these studies~\cite{2003-thieme-fy}. 
 
In contrast, in standard bulk-culture bacterial growth studies, the dynamical range in the population number of cells can be readily made to span multiple logarithmic decades in a day or two, since the typical timescales involved are of the order of minutes. Thus precise estimation of the exponential growth rate, $k$, defined to be the inverse of the time taken for the numbers of cells to increase by a factor of $e$, is feasible in this context~\cite{2010-hagen-pb}. Since many kinds of cells divide precisely into two cells at each division, the bulk growth rate is often used to infer the cell doubling time, i.e., the time it takes for a {\em single} cell to fission into two daughter cells, using the formula $(\ln 2)/ k$~\cite{2010-hagen-pb}. However, the extrapolation from the observed growth rate of the population (typically involving $\mathcal{O}(10^{8})$ cells per ml) to the inferred dynamics at the single-cell level is often inaccurate because all cells do not divide at precisely the same time after birth. There is significant stochasticity in the division times, as previously noted. One must account for this variability in relating the stochastic single-cell division dynamics to the population growth; even the mean population growth rate depends on the {\em shape} of the division time distribution.

A technical challenge in developing an exact theoretical framework for predicting population level behaviors, consistent with underlying stochastic single-cell dynamics, is that the aging dynamics of individual cells are non-Markovian or history dependent. This is reflected in the non-exponential shape of typical inter-division time distributions (waiting-time distributions for duration between successive divisions). Since standard techniques for finding exact solutions to stochastic processes are only applicable to Markovian or memoryless dynamics, we have used first principles to derive a non-Markovian description of the cell division process, including exact analytical time-dependent  solutions. These results represent a significant advance in the general theory of non-Markovian dynamics.

A unique advantage of studying balanced exponential growth in the microbial context is that high quality data are accessible at multiple scales of observation (sub-cellular, organismal and population level). In particular, the technology that we have recently developed for {\em C. crescentus} cells has the advantage that isolated single-cells in highly reproducible and unlimiting balanced-growth conditions can be observed with unprecedented statistical precision~\cite{2014-iyer-biswas-mz}; complementary approaches have been introduced by others for other organisms. Thus direct comparison between these single-cell experiments and bulk-culture measurements under the same growth conditions is possible.

\begin{figure}[t]
\begin{center}
\includegraphics[width= 1.\columnwidth]{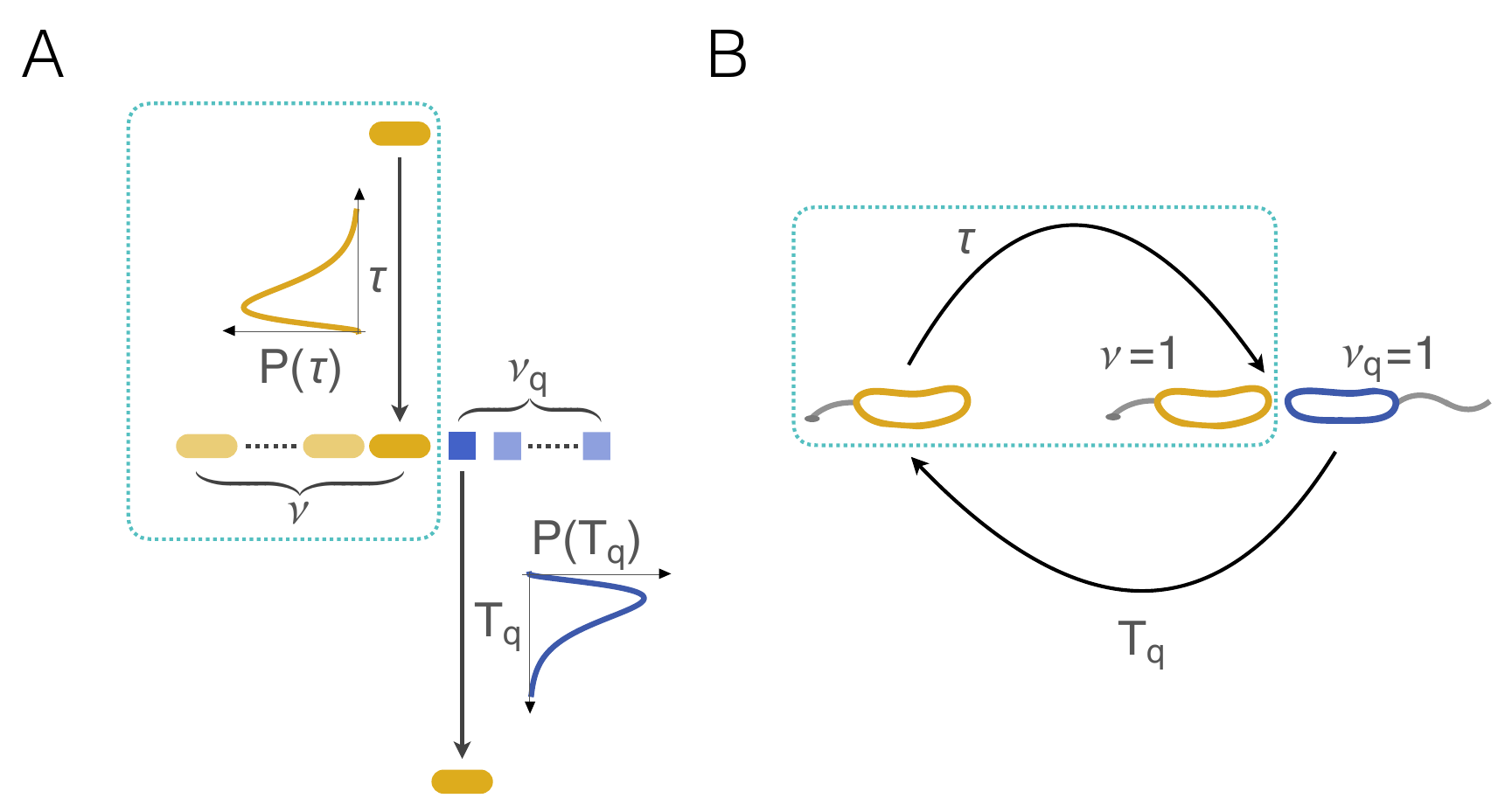}
\caption{{\bf Schematic representation of model and experimental system.} In (A) we summarize the general model considered. Upon division, a normal (yellow) cell divides into $\n$ normal cells and $\n_{q}$ reproductively quiescent (blue) cells. Division occurs after a stochastic waiting time $\t$, drawn from a  distribution, $P(\t)$. The quiescent cells transition into normal cells with a waiting-time distribution, $P_{q}(T_{q})$. Model specification requires knowledge of $\n, \n_{q}, P(\t)$ and $P_{q}(T_{q})$. By setting $\n_{q} =0$ one obtains a model of symmetric division (dotted cyan box). In (B) we show a schematic of the relevant timescales in the {\em C. crescentus} lifecycle: the stalked (normal) cell division time, $\t$, and the swarmer (quiescent) cell to stalked cell transition time, $T_{q}$. Upon division a stalked cell gives birth to a stalked and a swarmer cell. Thus the {\em C. crescentus} cell-cycle reduces to the model in (A) with $\n = \n_{q} = 1$. The single-cell technology introduced in \cite{2014-iyer-biswas-mz} corresponds to a model with $\n=1$ and $\n_{q}=0$ (dotted cyan box).}
\label{fig-schematic}
\end{center}
\end{figure}

A special feature of {\em C. crescentus} cells is that they divide asymmetrically. A replication-competent stalked cell (referred to here as a normal cell) grows and divides into two distinct daughter cells:  another stalked cell and a reproductively quiescent swarmer cell, which must undergo an additional differentiation step before transitioning into a stalked cell. The division and differentiation steps are thus controlled by distinct stochastic waiting-time distributions (see Fig.~\ref{fig-schematic}). Therefore, we generalize the theory to include asymmetric division.

It has been challenging to directly quantify the quiescence timescale, i.e., the time for a swarmer to differentiate into a stalked cell (labeled $T_{q}$ in Fig.~\ref{fig-schematic}). The single-cell technology has allowed direct visualization of growth and division dynamics of stalked cells~\cite{2014-iyer-biswas-mz}, but growth and division dynamics of individual swarmer cells are yet to be directly observed. Indeed, it is unknown how individual swarmer cells grow between division and differentiation, i.e., whether they grow exponentially in size, like stalked cells, or if there is a rapid growth spurt at a specific phase of differentiation. Population growth measurements typically involve mixed populations of stalked and swarmer cells and obscure details of swarmer cell dynamics. Knowledge of the quiescence timescale is essential for any quantification involving the full cell cycle of the organism. We address this gap in understanding by using the theoretical machinery developed herein to provide a prescription for estimating the quiescence timescale. For  {\em C. crescentus} cells, we also show that a simple measurement of the population growth rate, for a given growth condition, is sufficient to characterize the condition-specific cellular unit of time, and thus yields the mean (single-cell) growth and division timescales, fluctuations in cell division times, the cell age distribution, and the quiescence timescale.

\section{Results}

{\bf System characterization and notation.}
Since the number of cells per ml in typical bulk-culture measurements is very large (e.g., $\mathcal{O}(10^{8})$), we can treat the total number of cells in the population at a time $t$, ${N}(t)$, as a continuous variable. Moreover, it is reasonable to assume that the fluctuations in ${N}(t)$ relative to its mean value are negligible. In asymptotic balanced growth, ${N}(t)$ is expected to grow exponentially, i.e.,  ${N}(t) = {N}(0) \exp{(k\, t)}$.

Here, $k$ denotes the exponential growth rate of population size; it can be experimentally obtained from standard bulk-culture measurements. We denote the age of the cell, i.e., the time since the last division, by $\t$; $P(\t) d\t$ is the probability that a cell lives to age $\t$ and then divides between ages $\t$ and $\t + d \t$. In contrast,  the division {\em propensity}, $\a(\t)$, is defined as follows:  $\a(\t) d\t$ is the probability that a given cell of age $\t$  divides between $\t$ and $\t +d \t$. One might be tempted to equate it to $P(\t)$, but the two are distinct and related as follows: $\a(\t) =  {P(\t)}/{\le[1- \int_{0}^{\t} d\t' P(\t')\ri]}$ (see the Appendix for details). We note that in balanced growth, the division time distribution, $P(\t)$, is independent of the time of observation, $t$. 

In microbial systems, the number of progeny per cell is typically a constant number (a positive integer, such as 1 or 2); we denote it by $\n$. We define the age-dependent number density, $n(t, \t)$, as follows: $n(t, \t) d\t$ is the number of cells present at $t$, with ages between $\t$ and $\t+ d\t$. In this paper we consider growth conditions in which the probability of cell-mortality is negligible. Thus the total number of cells (and so the number of cells of each age) increases indefinitely. Mathematically, being in the balanced growth state means that the {\em fraction} of cells of each age at any time, $t$, is time invariant. Therefore, the cell-age distribution, $G(t, \t) \equiv n(t, \t)/ {N}(t)$, is time-independent in the long time (balanced growth) limit. We denote this steady-state age distribution by $G^{*}(\t)$. It is a normalized probability density since the sum of fractions of cells at each age is unity. In practice, single-cell measurements yield $P(\t)$, from which the division propensity, $\a(\t)$, can be computed (see above). Thus, the question then becomes how, given $P(\t)$ and $\n$,  the corresponding  {population} exponential growth rate, $k$, and the age distribution, $G^{*}(\t)$, are to be self-consistently determined.

{\bf General solution for symmetric cell-division.} To place the general solution in context, it is useful to first consider two familiar limiting cases of the problem.
(i) In the {\em deterministic} limit, cells divide exactly at age $\t_{o}$, i.e, $P(\t) = \d_{\t, \t_{o}}$. Therefore, ${N}(t) = {N}(0) \, \n^{(t/\t_{o})}$. Here, when $\n =2$, the doubling time is equal to the division time, $\t_{o}$. (ii) When the the dynamics are {\em Markovian} or memoryless, the division time distribution is  an exponential, $P(\t) =  \k \, e^{-\k \t}$. The dynamics in this case are identical to those of a one-step stochastic Hinshelwood cycle~\cite{2014-iyer-biswas-jk}, or equivalently, an age-dependent Galton-Watson process with an exponential waiting-time distribution~\cite{2008-feller-hc}. The full solution is known; in particular, ${N}(t) = {N}(0)\,e^{\k t}$. Thus the population growth rate, $k$, is equal to the single-cell exponential waiting time distribution parameter, $\k$. Note that even for this simple case, the mean division time of single cells, $\m_{\t} = 1/k$, is not equal to the mean doubling time of the population, $(\ln  2)/k$.

In general, the propensity of a cell to divide depends on its age, i.e., $\a$  varies with $\t$.  Consequently,  the time evolution is non-Markovian, and  $P(\t)$ is non-exponential. This significantly increases the complexity of the problem of finding the population growth rate, $k$, as a functional of $P(\t)$; for symmetric division, the process falls in the category of a Bellman-Harris branching process~\cite{2008-feller-hc}. Typically, the division time distribution is a unimodal distribution with a peak at a finite time and a positive skew, i.e., it has a long right tail. Thus, it is qualitatively different from the monotonically decreasing exponential distribution (Markovian limit). Therefore solving the general non-Markovian case is important. We derive the time evolution equations for the age-dependent number density, $n(t, \t)$, and the age distribution, $G(t, \t)$, for a general $P(\t)$ and solve them exactly (see Appendix for details). 

In the general case, the population's exponential growth rate, $k$, is related to the single-cell division time distribution, $P(\t)$, through the integral
\begin{align}
\label{eq-k-soln}
\le\la e^{-k \t} \ri\ra_{P}  \equiv \int_{0}^{\infty} d\t\, e^{-k \t} P(\t)  = {\frac{1}{\n}}. 
\end{align}
In words, $k$ is the point at which the Laplace transform of the division time distribution, $P(\t)$, is equal to $1/\n$. The  expressions for the age distribution and the total population number, for a given initial condition, ${N}(0)$, are
\begin{align}
\label{eq-age-soln}
&G^{*}(\t) =  \frac{\n\,k} {(\n-1)}\, e^{ - k \t} \le[ 1 - \int_{0}^{\t} d\t' \, P(\t') \ri]; \nn \\
&{N}(t) = {N}(0)\, e^{k \, t}; \nn \\
&n(t, \t) \equiv {N}(t)\,G^{*}(\t)
= {N}(0) \frac{ \n k } {(\n-1)}\, e^{ -\le[ \int_{0}^{\t} d\t' \a(\t')\ri]}.
\end{align}
Together, Eqs.~\eqref{eq-k-soln} and \eqref{eq-age-soln} constitute the complete analytical solution to the problem for symmetric cell division.

From the general solution it follows that there is a unique steady-state age distribution, $G^{*}(\t)$, corresponding to a given division time distribution, $P(\t)$, and progeny number, $\n$. Conversely, if the (population) cell-age distribution and the bulk exponential growth rate are observed, Eq.~\eqref{eq-age-soln} can be used to infer the {\em single-cell} division time distribution! We note that the growth rate, $k$, is itself a functional of $P(\t)$ for a given $\n$, and is thus not an independent parameter of the solution for $G^{*}(\t)$ in Eq.~\eqref{eq-age-soln}. 

Since the shape of the age distribution reveals features of the division time distribution, its qualitative features are of interest. Briefly, they are as follows. (See Fig.~S1 for a graphical summary of these results.) First, the age distribution monotonically decreases with $\t$. To see this, note that in Eq.~\eqref{eq-age-soln},  the cumulative integral $\int_{0}^{\t} d\t' \, P(\t')$ increases with $\t$ since $P(\t)>0$. Next, since $P(\t) \propto d \,(e^{k \t} G^{*})/d \t$, the most probable division time is determined by where the curvature of $e^{k \t} G^{*}(\t)$ changes sign, i.e., its point of inflection. Also note that the slope of this function at its point of inflection estimates the width of the division time distribution. For $\n = 1$, a case considered in detail below, the mean division time is given by the point of inflection of the age distribution and the slope at this point estimates the width of the division time distribution. Finally,  for a given $P(\t)$, a greater value of $\n$ (number of progeny per cell) will increase the growth rate and skew the age distribution towards smaller ages (i.e., to the left).   

{\bf Comparison with single-cell experiments: measured and predicted cell-age distributions.} We compare our theory to recent single-cell data for {\em C.\ crescentus} \cite{2014-iyer-biswas-mz}.  {\em C.\ crescentus} divides into two morphologically and functionally distinct daughter cells: an adherent stalked cell that is replication competent and a motile swarmer cell that cannot divide further but can differentiate into a stalked cell.  In these microfluidic experiments \cite{2014-iyer-biswas-mz}, stalked cells are retained and  swarmer cells are removed after each division (see Fig.~\ref{fig-schematic}). Therefore the stalked-cell dynamics (in these experimental conditions) are equivalent to cells being simply ``renewed'' after each division, i.e.,  $\n =1$. Also, the total number of stalked cells in the experiment, ${N}(t)$, is constant. These features simplify the problem, and we can use the analytical results for the symmetric-division model. However, the growth dynamics are still non-Markovian and hence non-trivial. For $\n=1$ the relation between the division time and age distributions becomes
\begin{align}
\label{eq-nu1}
G^{*}(\t) = \frac{1}{\m_{\t}}\le[1 - \int_{0}^{\t} d\t' P(\t')\ri],
\end{align}
where ${\m_{\t}}$ is the mean division time, $\m_{\t} \equiv \int_{0}^{\infty} d\t\, \t P(\t)$. 
See the Appendix for details and Fig.~S1 for a graphical interpretation.

\begin{figure}[ht]
\begin{center}
\includegraphics[width= 1.\columnwidth]{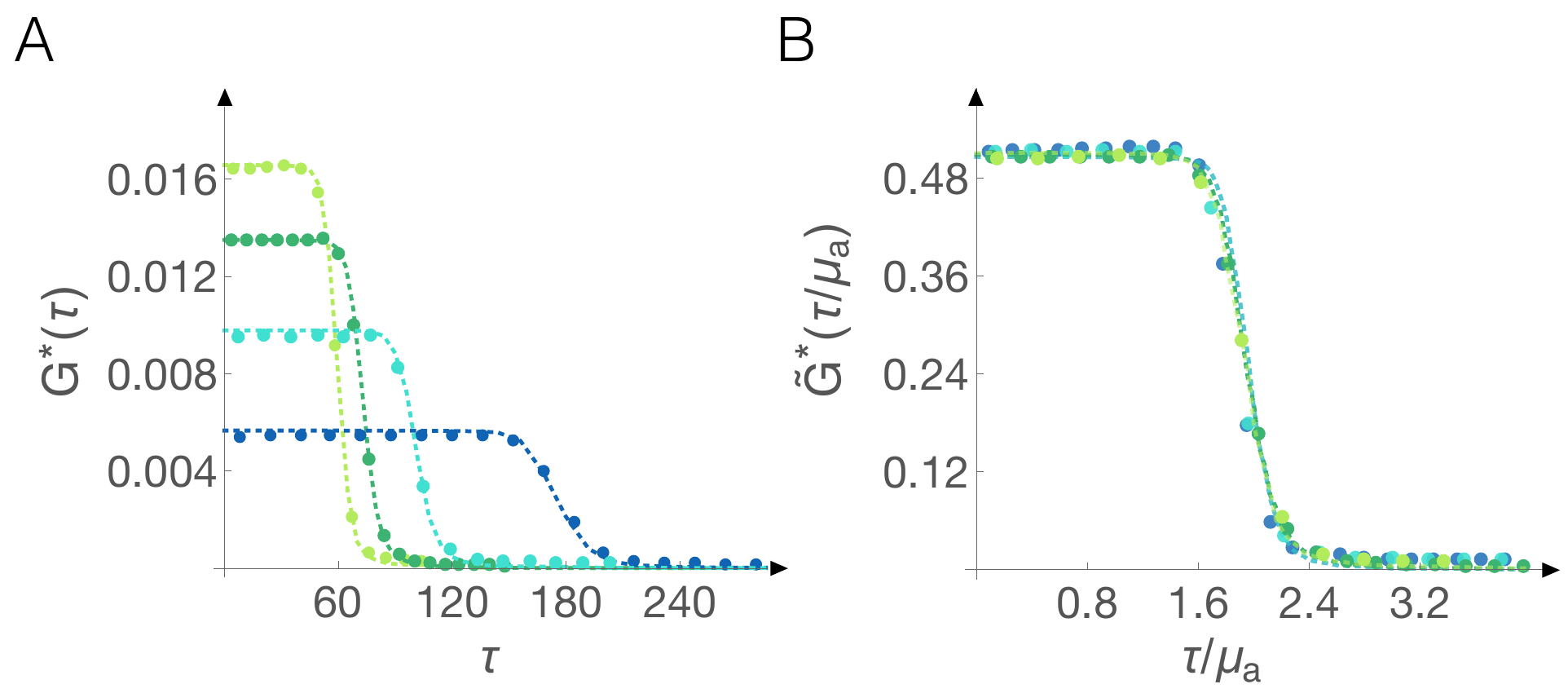}
\caption{{\bf Cell age distributions, measured and predicted, from different temperatures.} In (A) we show the age distributions from different temperatures (light green, 34$^{\circ}$C; dark green, 31$^{\circ}$C; cyan, 24$^{\circ}$C; and blue, 17$^{\circ}$C). The data from single-cell experiments are shown with circular symbols and the predictions from the theory (with no adjustable parameters) are shown with corresponding dashed lines. In (B) we show the same age distributions as in (A), rescaled by their respective temperature dependent mean ages. $\tilde{G}^{*}(\t/\m_{a}) \equiv \m_{a} G^{*}(\t)$ is the mean-rescaled probability density of cell ages. Evidently, once mean-rescaled, the probability densities undergo a scaling collapse, consistent with a single condition-specific timescale dominating stochastic growth and division statistics (see text). For the measured cell division time distributions corresponding to these age distributions see Fig.~3B of ~\cite{2014-iyer-biswas-mz}.}
\label{fig-age-dist}
\end{center}
\end{figure}

We use the measured time-courses of single-cell growth and division to validate the theory as follows.  We obtain the {\em measured} cell-age distribution  by building a histogram of the duration between the time of observation and the recorded time of the previous division for each cell. From the same  experiment,  we also measure the division time distribution, and we insert it into Eq.~\eqref{eq-nu1} to compute the {\em predicted} age distribution. For experiments spanning the physiological temperature range of the organism, we find excellent agreement between the measured and predicted cell-age distributions with no fitting parameters (Fig.~\ref{fig-age-dist}A), confirming the model.

When rescaled by their condition-specific means, cell-age distributions from all temperatures collapse onto a single curve (Fig.~\ref{fig-age-dist}B symbols). The underlying physical principle encoded in this universal behavior is that a single temperature-dependent scale of time, proportional to the mean division time (or equivalently, the mean age), governs stochastic growth and division dynamics \cite{2014-iyer-biswas-jk, 2014-iyer-biswas-mz}. For the scaling collapse of the corresponding mean-rescaled division time distributions see~\cite{2014-iyer-biswas-mz}. 

We note that the mean age, $\m_{\tiny{a}}$, and mean division time, $\m_{\t}$, are not equal. For $\n=1$, using Eq.~\eqref{eq-nu1}, the two are related by
\begin{align}
\label{eq-means}
\m_{\tiny{a}} = \frac{\m_{\t}}{2}\le(1 + \eta_{\t}^{2} \ri),
\end{align}
where $\eta_{\t}$ is the COV of the $P(\t)$ distribution. (See Appendix, Eq.~\eqref{eq-start} to Eq.~\eqref{eq-stop}, for a detailed derivation.) Using Eqs.~\eqref{eq-nu1} and \eqref{eq-means}, we find the predicted mean-rescaled age distribution; this also agrees excellently with the observed distribution (Fig.~\ref{fig-age-dist}B line). Eq.~\eqref{eq-means} shows that when there is no stochasticity in division times (i.e., the deterministic case $P(\t)=\delta_{\t,\t_o}$), then the mean age is half the mean division time since cell age  is uniformly distributed from $0$ to $\t$. Interestingly, this provides a lower bound for the mean age (for a specified mean division time) since any stochasticity in $P(\t)$ can only increase $\eta_{\t}$, and thus the mean age. We note that in practice, even if there is sizable noise in division times, the second term is negligible compared to the first (for $20\%$ noise in division times, the ratio of the two terms is $0.04$). Thus, $\m_{\tiny{a}} \approx \m_{\t}/2$.

\begin{figure}[t]
\begin{center}
\includegraphics[width= 1.\columnwidth]{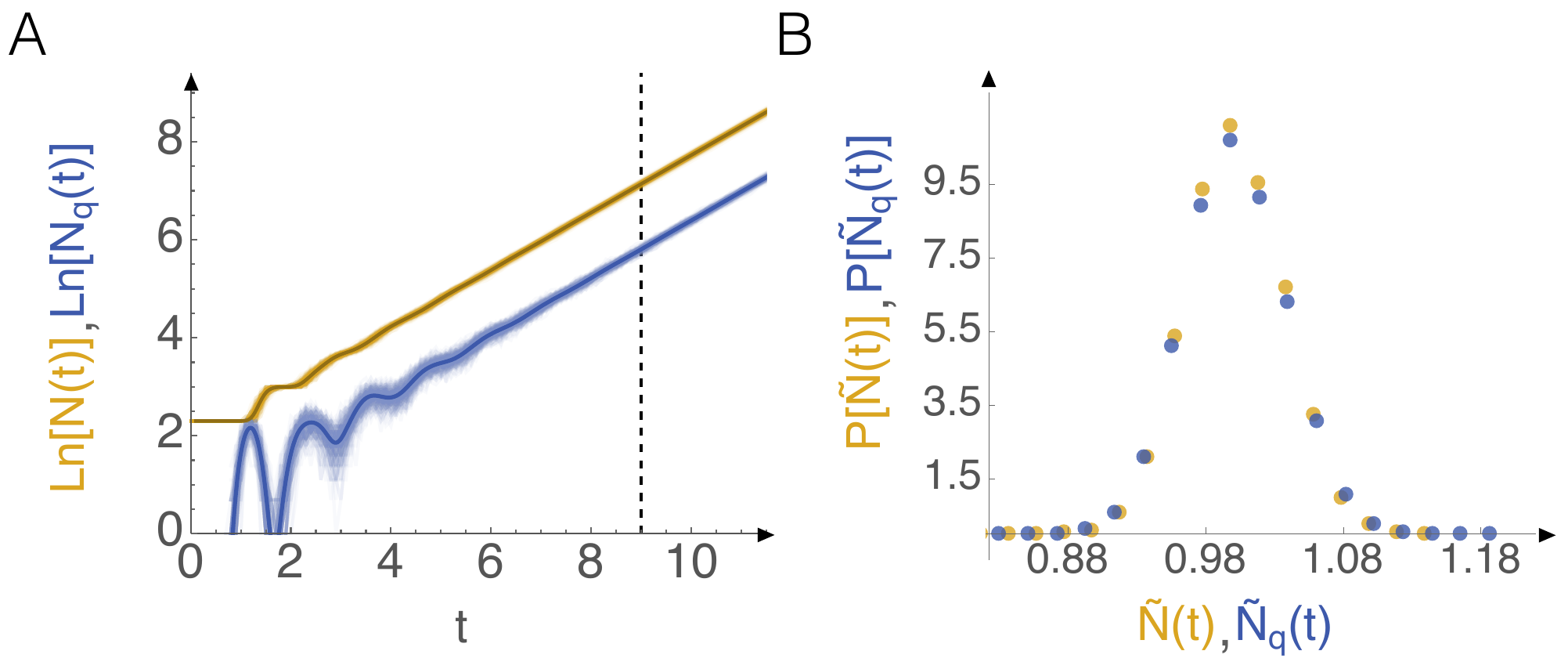}
\caption{ {\bf Scaling of number fluctuations during stochastic population growth.} In (A), we show analytical results for the time evolution of population sizes of normal (yellow) and quiescent (blue) cells. In steady-state, they have the same exponential growth rate.  Oscillations in numbers of normal and quiescent cells are observed during the transient phase. The bold curves show the mean behavior of population numbers, calculated using our exact analytical solution. Also shown are representative stochastic trajectories which account for fluctuations in population numbers. The stochastic trajectories were computed using exact numerical simulations (see accompanying text).  In (B) we show that the mean-rescaled population number distributions of normal and quiescent cells (from $t=9$ in panel (A)) undergo a scaling collapse. Here, $\tilde{N}(t) = N(t)/\la N(t)\ra$ and $\tilde{N_{q}}(t) = N_{q}(t)/\la N_{q}(t)\ra$. The parameter values used for these simulations are: $\n = \n_{q} = 1$; $N(0) = 10$; $N_{q}(0)=0$; $P(\t)$ is a Gamma distribution with mean = 1 and COV = 0.13 and $P_{q}(T_{q})$ is a Gamma distribution with mean = 0.4 and COV = 0.1. All timescales are measured in units of $\m_{\t}$.}
\label{fig-num}
\end{center}
\end{figure}

Once the age distribution (for a specific balanced growth condition) has been determined, it can be used to deconvolve cell-cycle phase dependence from a population of asynchronous cells, since it predicts the probability weight to associate with cells of each age. This obviates the need for less precise bulk-synchronization experiments, in which it is also unclear how the synchronization procedure may itself alter the balanced growth state. For instance, if the initial population has only swarmer cells, then the numbers of stalked and swarmer cells in the population oscillate with time and the culture is far from being in balanced growth (Fig.~\ref{fig-num}). We note that an early empirical algorithm for cell-cycle phase deconvolution was given in~\cite{2009-siegal-gaskins-ec}.

{\bf Generalization to asymmetric division.} Motivated in part by the {\em C.\ crescentus} data discussed above, we now generalize the theory to  allow for asymmetric divisions (see Fig.~\ref{fig-schematic}). There are two distinct cell types in the population:  normal division-capable cells, and  reproductively quiescent cells which take an additional stochastic waiting time, $T_{q}$,  to differentiate (transition) to normal cells before being able to divide.  Each normal cell divides into $\n$ normal cells and $\n_{q}$ quiescent cells. The waiting time $T_q$ has  probability distribution $P_{q}(T_{q})$.  Normal cells divide with a division time distribution, $P(\t)$, as before. With the inclusion of asymmetric division, the process is no longer a standard branching process, since different cells in the population are not statistically identical~\cite{vankampen}. This significantly increases the complexity of the problem. However, we have found exact analytical solutions (see the Appendix for details). We denote the (steady-state) age distributions of normal and quiescent cells by $G^{{*}}{(\t)}$ and $G^{{*}}_{q}(T_{q})$, respectively. In the $\n_{q}=0$ limit the problem becomes equivalent to the symmetric division case.

A key physical insight is that the ratio of normal to quiescent cells should be a constant for balanced growth conditions. Consequently, both kinds of cells must increase exponentially in numbers, with the {\em same} growth rate, $k$. See Appendix for details. The exact solution for $k$, for specified functional forms of $P(\t)$ and  $P_{q}(T_{q})$, when $\n$ normal and $\n_{q}$ quiescent progeny are born at each division, is
\begin{align}
\label{eq-sw-st}
\le\la e^{-k \t} \ri\ra_{P} = \frac{1}{\n + \n_{q}\, \le\la e^{-k T_{q}} \ri\ra_{P_{q}}}.
\end{align}
In this solution $ \le\la e^{-k \t} \ri\ra_{P} \equiv \int_{0}^{\infty} d\t P(\t) e^{-k \t}$ and $\le\la e^{-k T_{q}} \ri\ra_{P_{q}} \equiv \int_{0}^{\infty} d T_{q} \,P_{q}(T_{q}) e^{-k T_{q}}$. For the complete solution, including expressions for $G^{{*}}{(\t)}$, $G^{{*}}_{q}(T_{q})$, and the fixed ratio of stalked to swarmer cells, see Appendix.

{\bf Comparison with population level experiments: scaling of timescales in the {\em C. crescentus} lifecycle.}
For {\em C. crescentus} cells, the ``normal'' cells correspond to stalked cells, and ``quiescent'' cells correspond to  swarmer cells. Each stalked cell divides into a stalked cell and a swarmer cell. The swarmer differentiates into a stalked cell after a time, $T_{q}$. Thus, for bulk-culture experiments with {\em C. crescentus} cells, $\n = \n_{q} = 1$, and Eq.\ \eqref{eq-sw-st} becomes
\begin{align}
\label{eq-sw-st-CC}
\le\la e^{-k \t} \ri\ra_{P} = \frac{1}{1 + \, \le\la e^{-k T_{q}} \ri\ra_{P_{q}}}.
\end{align}
From our bulk-culture experiments, we are able to determine the population growth rate, $k$. Population growth data were obtained using standard optical density measurements. For each experimental condition $12$--$20$ growth curves were obtained under dilute growth conditions; for each growth curve we recorded $6$--$10$ data points in the ``log--phase''. The exponential growth rate of the population was determined by averaging the growth rates obtained from the log phase data of each growth curve. Moreover, $P(\t)$ is also known, since it is directly observed in our single-cell experiments~\cite{2014-iyer-biswas-mz}. Thus, the timescale that remains to be determined is the swarmer-to-stalked cell transition time, $T_{q}$, and the corresponding distribution, $P_{q}(T_{q})$. There are several technical reasons why direct experimental characterization of $T_{q}$ is challenging (see~\cite{2010-england-uk}). Yet knowledge of this timescale could provide an important clue to many fundamental biological questions. For example, it is not known precisely what fraction of the {\em C. crescentus} lifecycle is spent in the swarmer stage. Determining this may in turn indicate how the additional differentiation step in the lifecycle confers flexibility to the fitness of {\em C. crescentus} cells in different growth conditions. Also, results in~\cite{2014-iyer-biswas-mz} imply that there must be cell size growth at the swarmer stage, since the average size of a newborn swarmer is only $\approx 80\%$ of the average size of a newborn stalked cell, and the newborn stalked cell size distribution has been shown to be invariant. But whether swarmer cell sizes increase linearly, exponentially, or in a rapid growth spurt during differentiation, remains to be determined.

\begin{figure}[ht]
\begin{center}
\includegraphics[width= 1.\columnwidth]{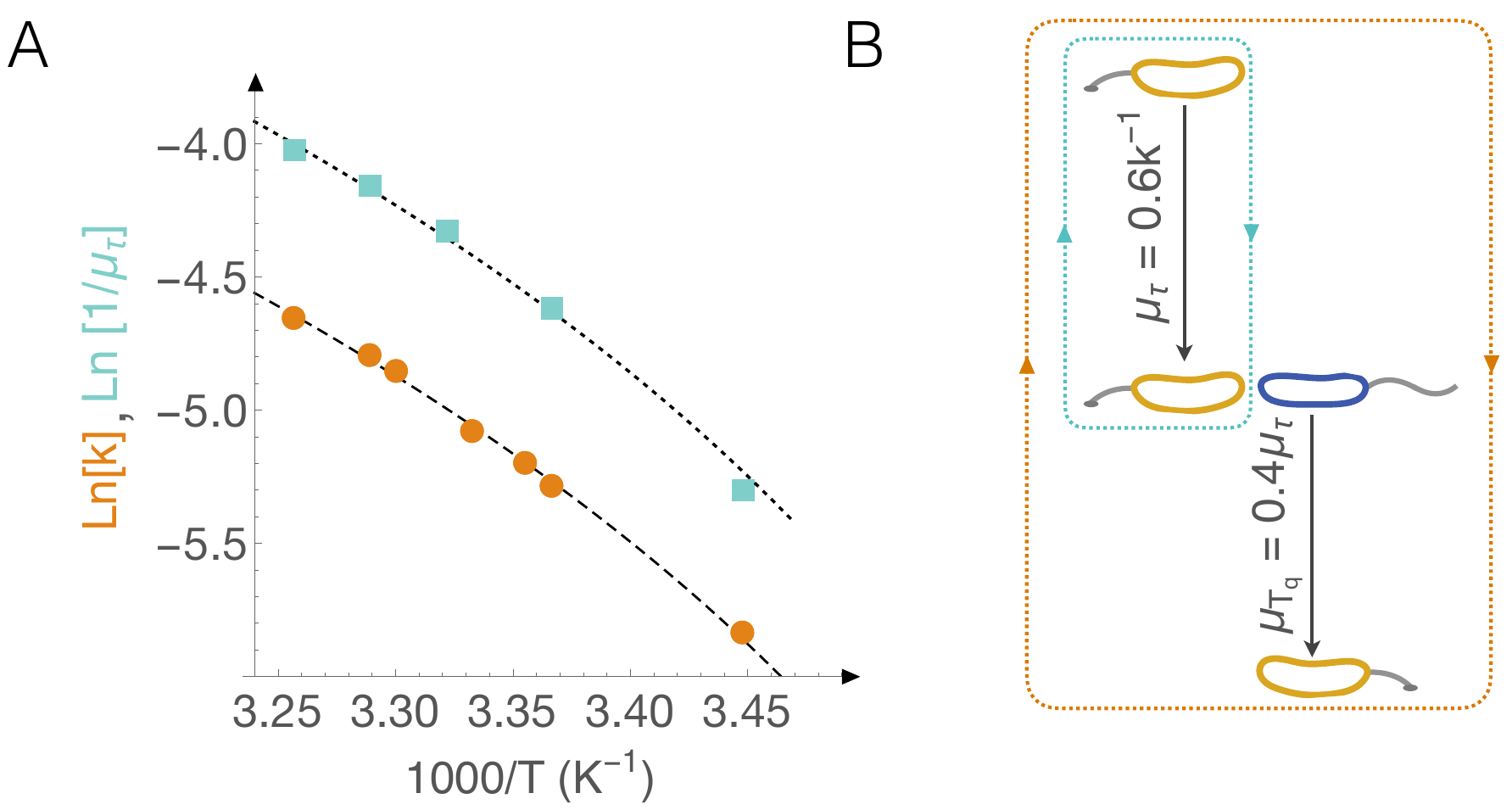}
\caption{{\bf Scaling of timescales in the {\em C. crescentus} lifecycle.} (A) The population growth rate, $k$ (orange circles), and the single-cell mean division rate from\cite{2014-iyer-biswas-mz} (cyan squares) are shown  as functions of temperature on an Arrhenius plot. The measurement precision is better than the sizes of the plot markers. The dotted (dashed) line is the best fit Ratkowsky curve~\cite{2014-iyer-biswas-mz}  for the single-cell (population) data. Both fits are found to be proportional to each other, with Ratkowsky temperature 269 K; by finding the ratio of the two curves, we find that $\m_{\t} k = 0.6$. (B) Combining the results from (A) with Eq.~\eqref{eq-sw-st-CC} and results from \cite{2014-iyer-biswas-mz}, we are able to characterize all timescales in the {\em C. crescentus} lifecycle as fractions of mean single-cell division time, or equivalently, the inverse of the population growth rate (see accompanying text). At different temperatures all timescales change proportionally, the proportions are shown in (B). Single-cell measurements sample the dynamics of stalked cell division (cyan rectangle), whereas population growth measurements sample the dynamics of stalked cell division and swarmer cell differentiation (orange rectangle).}
\label{fig-timescales}
\end{center}
\end{figure}

Here we estimate the swarmer-to-stalked transition time for different balanced growth conditions using Eq.~\eqref{eq-sw-st-CC}. The results are shown in Fig.~\ref{fig-timescales}. At each growth condition, we invert the (integral) equation to estimate $T_{q}$, using the experimentally determined $k$ (the bulk growth rate) and $P(\t)$ (the stalked cell division time distribution). The {mean} value of this timescale, $\m_{T_{q}}$, is insensitive to the particular functional form assumed for $P_{q}(T_{q})$. Therefore, we use $P_{q}(T_{q}) = \d(T_{q} -\m_{T_{q}})$ and find $\m_{\tiny{T_{q}}}$ using Eq.~\eqref{eq-sw-st}, after numerically evaluating the Laplace transform for $P(\t)$ at $k$ for each growth condition. Remarkably, the {\em fraction} of the cell cycle spent in the swarmer stage is a constant, as temperature is varied  (and the duration of the lifecycle itself changes by a factor of $\approx 4$). Specifically, we find that {$ \m_{\t} k = 0.6$ and $\m_{T_{q}} = 0.4 \m_{\t}$} (Fig.~\ref{fig-timescales}). This result is consistent with indirect measurements in~\cite{2010-england-uk}. Moreover, using this ratio, we can predict the ratio of swarmer to stalked cells in the population during balanced growth (see Appendix for details). We find that ${N}_{q}(t)/{N}(t) \approx 0.2$, also consistent with previous estimates~\cite{2010-england-uk}, further validating our approach. 

In \cite{2014-iyer-biswas-mz} we showed that the {\em single-cell} exponential growth rate (of stalked cell sizes), $k_{\tiny{sc}}$, determines a condition-specific cellular unit of time. Thus, it governs all aspects of the stochastic dynamics of stalked-cell growth and division; in particular, its inverse is proportional to the mean division time, $\m_{\t}$, and it also determines the full distribution, $P(\t)$. Since we now find that $\m_{T_{q}}/\m_{\t}$ is also a constant, the implication is that the single-cell exponential growth time scale, $k_{\tiny{sc}}^{-1}$, proportional to the population growth rate, $k$, governs {all} relevant timescales for growth, division {\em and differentiation}.  Thus {\em all} timescales rescale proportionally, when external conditions are changed (see Fig.~\ref{fig-timescales}). The remarkable implication is that for any balanced growth condition of interest, a simple measurement of the population growth rate, $k$, when used to rescale the universal mean-rescaled distributions we have found (Fig.~\ref{fig-age-dist} and \cite{2014-iyer-biswas-mz}), together with the model, yields distributions of cell-division times, cell ages, and cell sizes.

Moreover, we find that the population and single-cell exponential growth rates are approximately equal to each other for all temperatures in Fig.~\ref{fig-timescales}: $k_{\tiny{sc}} \approx k$.  The surprising implication of this observation is that the duration of the swarmer-to-stalked cell transition time is accounted for by the fact that cell numbers {\em double} in the time that cell sizes increase by a factor of {$1.8$}. This observation is consistent with swarmer cell sizes also increasing exponentially with time, with the same growth rate. However, validation of this interpretation requires further experimentation. 

\begin{figure}[t]
\begin{center}
\includegraphics[width= 1.\columnwidth]{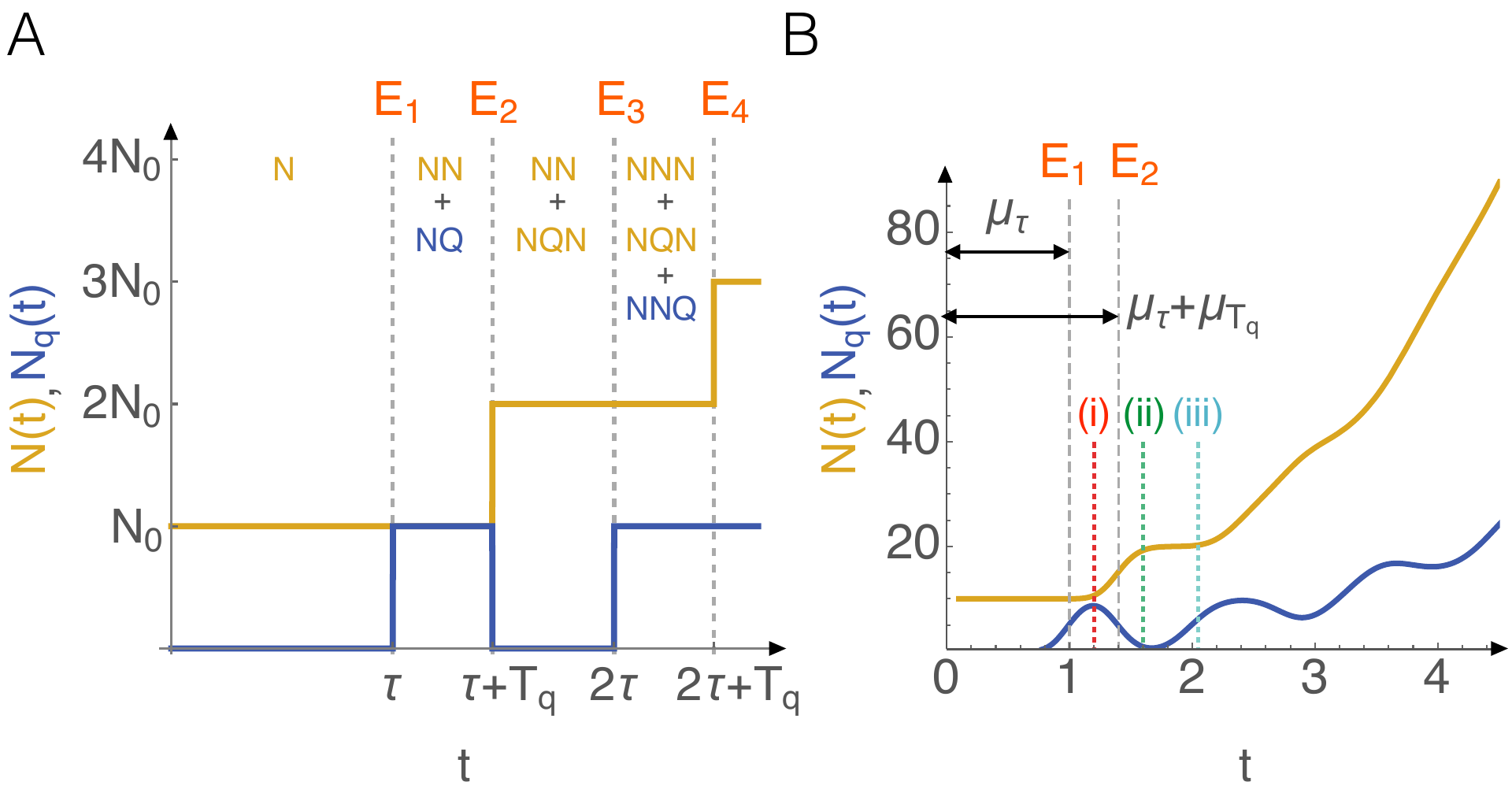}
\caption{ {\bf The physics of transient oscillations.} In (A) we sketch a simple example that illustrates how transient oscillations in population numbers arise. See accompanying text for notation and definition of variables. Initially, $N_{0}$ normal cells are present, and population numbers increase as these cells divide and differentiate. The possible lineages of cells present in each demarcated interval are shown as the appropriate sequence of normal (N) and quiescent (Q) stages in their ancestry, starting at $t=0$, and including their current state. At $t=\t$, all cells of lineage `N' divide, causing an increase by $N_{0}$ in $N_{q}(t)$, while $N(t)$ remains equal to $N_{0}$. At $t=\t + T_{q}$, the quiescent cells differentiate synchronously, creating the lineage `NQN'. This results in a simultaneous decrease in $N_{q}(t)$ and increase in $N(t)$, each by $N_{0}$. The next change in numbers occurs with the division of all cells of `NN' lineage at $t=2\t$. In this manner the persistence of distinct lineage identities causes distinct fractions of the population to undergo division or differentiation synchronously, resulting in the observed population oscillations. In (B) we show the corresponding transient oscillations in population numbers for a realistic model where division and differentiation occur probabilistically. See Fig.~\ref{fig-num} legend for parameter values used; for these parameter values we find that $\m_{\t} = 1$ and $\m_{T_{q}} = 0.4$.  From the times at which the first oscillation in quiescent cell populations begins and ends, we can deduce the mean cell division ($\m_{\t}$) and differentiation ($\m_{T_{q}}$) times (compare with panel (A)). The age distributions at the transient times (i)--(iii), corresponding to stages $2-4$ in (A), are shown in Fig.~\ref{fig-td-age}.  All curves were calculated from our exact analytical time-dependent solutions (see the Appendix for details).}
\label{fig-osc}
\end{center}
\end{figure}

{\bf Transient dynamics and oscillations in population numbers.} 
To extend the results to time-dependent scenarios, i.e., to account for transient behaviors before steady state is attained, we have derived the exact analytical time-dependent solution to the general problem of asymmetric division (see the Appendix for derivation and analytical expressions).  Steady state solutions and results for symmetric division are obtained as straightforward limits of the general solution.

Surprisingly, in the transient regime, population numbers of both normal and quiescent cells may oscillate with time. As previously noted, in steady-state the numbers of normal and quiescent cells grow exponentially, with the same exponential growth rate. However, in the short time limit, depending on initial conditions, numbers of normal and quiescent cells may oscillate with time. See Figs~\ref{fig-osc}B and \ref{fig-num}A for typical instantiations of these oscillations.

We elucidate the physics of oscillatory transients through a simple example  (see Fig.~\ref{fig-osc}A). Consider deterministic division and differentiation: a normal cell divides into a normal and a quiescent cell after time $\t$, and the quiescent cell transitions into a normal cell after time $T_{q}$. In other words,  $P(\tau)$ and $P_q(T_{q})$ are delta functions peaked at times $\t$ and $T_q$ respectively, and $\n = \n_{q} = 1$. For brevity, let us assume that $\t > T_q$, and that initially $N=N_0$ normal cells and $N_q = 0$ quiescent cells are present. We denote the lineage of a cell at a given generation by the sequence of reproductively normal (N) and quiescent (Q) stages in its ancestry, beginning from $t=0$, and including its current stage. 

Nothing occurs until $t=\t$, when all cells of lineage `N' simultaneously divide, causing $N_q$ to increase by $N_0$ (event $E_{1}$ in Fig.~\ref{fig-osc}A). The next event occurs at $t=\t+T_q$, when all quiescent cells of lineage `NQ' differentiate into normal cells, causing $N_q$ to decrease to $0$, and $N$ to increase to $2N_0$ (event $E_{2}$ in Fig.~\ref{fig-osc}A). Following this event, the normal cell population consists of two subpopulations, each consisting of  $N_{0}$ cells of lineages `NN' and `NQN', respectively. The second subpopulation is $T_{q}$ younger in age than the first.  Next, at $t=2\t$ (event $E_{3}$ in Fig.~\ref{fig-osc}A), the older generation of normal cells, `NN', divide, causing $N_q$ to jump to $N_0$ again due to the birth of cells of lineage `NNQ'. Subsequently, at time $t=2 \t + T_q$, all $N_0$ normal cells from generation `NNQ' divide. At the same time, all quiescent cells of lineage `NNQ' differentiate into normal cells. The net result of these processes is that $N(t)$ jumps from $2N_0$ to $3N_0$. Thus periodic jumps and dips in population numbers continue to occur at times that are various integer combinations of $\t$ and $T_q$, resulting in oscillatory behavior of population numbers. In this manner, population oscillations arise when the initial population is highly synchronized, i.e., when the initial age distributions of normal and quiescent cells (if present) are narrow. The persistence of distinct lineage identities in subsequent generations, reflected in narrow age distributions of subpopulations grouped by lineages, causes distinct fractions of the population to undergo division or differentiation synchronously. This dynamic results in transient population oscillations.

\begin{figure}[ht]
\begin{center}
\includegraphics[width= 1.\columnwidth]{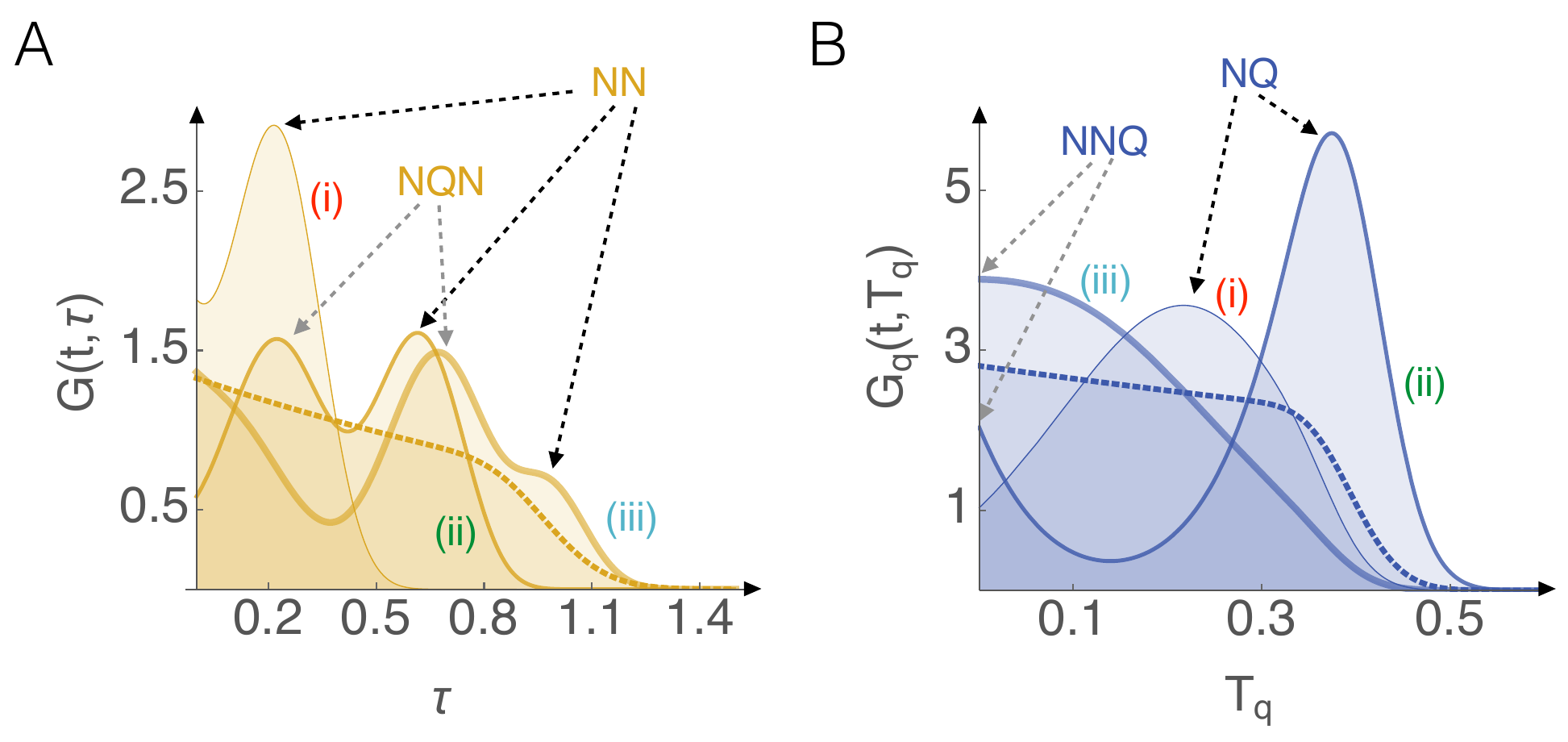}
\caption{ {\bf Time evolution of age distributions of normal and quiescent cells.} The solid curves in (A) and (B) show transient normal and quiescent cell age distributions, respectively. We have plotted these distributions at $t = 1.2, 1.6,$  and $2.05$, corresponding to times labeled (i), (ii), and (iii), respectively, in Fig.~\ref{fig-osc}B. See Fig.~\ref{fig-num} legend for parameter values used; for these parameter values we find that $\m_{\t} = 1$ and $\m_{T_{q}} = 0.4$. The initial population consists of newborn normal cells. Each mode of a time-dependent age distribution corresponds to a specific lineage, as labeled above. The separations between different modes of multimodal time-dependent age distributions contain information about underlying timescales of division and differentiation. For instance, in (A), curve (ii) shows a bimodal normal cell age distribution observed at $t=1.6$. The separation between the two modes (corresponding to the lineages `NN' and `NQN') is approximately equal to the differentiation timescale, $\m_{T_{q}} = 0.4$. The broadening of age distributions of each new lineage, due to fluctuations in division and differentiation times, results in steady-state age distributions (dashed curves) in which contributing lineages can no longer be distinguished. All curves were calculated from the exact analytical time-dependent solutions, which are presented in the Appendix. The $\d$-function corresponding to the initial age distribution has not been shown in (A). See Supplementary Videos 1 and 2 for the full time evolution of the age distributions corresponding to panels A and B, respectively.}
\label{fig-td-age}
\end{center}
\end{figure}
In a realistic model with finite widths for $P(\tau)$ and $P_q(T_{q})$, division and differentiation do not occur synchronously for the entire population. Thus, even when the initial population is perfectly synchronized, each subsequent event increasingly broadens the normal and quiescent cell age distributions. This desynchronization effect eventually wipes out population number oscillations, and the steady-state age distributions retain no signatures of lineage identities of subpopulations of cells. In the presence of stochasticity, population oscillations are no longer as sharply defined as in the deterministic case discussed previously (contrast the oscillations in Fig.~\ref{fig-osc}B with Fig.~\ref{fig-osc}A). Instead, the time distribution for each event is  given by the convolution of time distributions corresponding to the elementary processes (cell divisions and quiescent-normal transitions) leading up to that event. Consequently, later events desynchronize such that they are washed out in the population average. This is illustrated in Fig.~\ref{fig-osc}B, in which we have shown the time-dependent normal and quiescent cell population numbers, calculated using our exact analytical time-dependent solution. However, the first few events can still be distinguished in the oscillations, allowing estimation of the timescales corresponding to the elementary processes. For example, the first division and differentiation events can be identified in Fig.~\ref{fig-osc}B, as the rise and fall of the first bump in $N_{q}(t)$, yielding estimates of the mean timescales of cell division ($\m_{\t}$) and differentiation ($\m_{T_{q}}$) times. This provides a prescription for inferring underlying timescales of division and differentiation from transient population oscillations obtained by using initially synchronized populations.

The transition from oscillatory to smooth steady state behavior, due to the broadening of the age distributions of later generations, is also reflected in the temporal evolution of shapes of the normal and quiescent age distributions. This effect is evident in Fig.~\ref{fig-td-age}, where we have visualized the time evolution of the age distributions of normal and quiescent cells, calculated using the exact analytical time-dependent solution. Moreover, the separation between the different modes of the multimodal time-dependent age distributions contains information about the underlying timescales of division and differentiation. For instance, in Fig.~\ref{fig-td-age}A, curve (ii) shows a bimodal normal cell age distribution observed at $t=1.6$. The separation between the two modes (corresponding to the lineages `NN' and `NQN', as marked in the figure) is approximately equal to the differentiation timescale, $\m_{T_{q}} = 0.4$. Thus, the time evolution of age distributions of an initially synchronized population provides another route to infer the underlying timescales of division and differentiation.

{\bf Scaling of fluctuations in population numbers.} 
Since population growth studies typically involve very large numbers of cells, fluctuations in population numbers are negligible, and our analytical results average over these fluctuations. However, in experiments with small populations (at the intermediate scale between single-cell and population growth studies), these fluctuations may be non-negligible. We have used numerical simulations to extend analytical results and investigate these fluctuations. We performed stochastic simulations for population growth for different waiting-time distributions and progeny numbers for both symmetric and asymmetric division by developing an exact algorithm for simulating these non-Markovian population growth dynamics. We note that the standard Gillespie algorithm~\cite{gillespie} cannot be used, since it assumes exponential waiting-time distributions.

Our simulation results are summarized in Fig.~\ref{fig-num}. Remarkably, the fluctuations in population sizes show the following scaling behavior: once balanced growth is reached and transients die out, mean-rescaled distributions of numbers of normal and quiescent cells undergo a striking scaling collapse. Moreover, mean-rescaled distributions from different times (after the transients die out) are also identical. Thus, the shape of the population number distributions, in balanced growth state, is time-invariant. This scaling collapse is reminiscent of the scaling collapse observed for the general stochastic Hinshelwood cycle model~\cite{2014-iyer-biswas-jk}.  The phenomenological model of stochastic exponential growth that yields the observed scaling collapse is discussed in~
\cite{2017-pirjol-kx}.

\section{Concluding remarks}
In this paper we have introduced an exact theoretical framework for predicting {population} level behaviors which are {consistent} with underlying stochastic single-cell dynamics. Therefore, using this framework, future studies can use population level data to infer characteristics of stochasticity at the single-cell level (such as mean and variance of the single-cell inter-division time distributions), without making ad-hoc assumptions. This is a useful prescription, since population growth measurements are relatively straightforward even for systems for which single-cell dynamics are experimentally inaccessible. We have related single-cell division to population growth, by introducing an analytical framework that takes dynamics at both scales into account. We have validated this framework by matching predicted and observed age distributions for {\em C. crescentus} cells in balanced growth at different conditions. We have also used this framework to show how timescales characterizing these dynamics scale with external conditions (different temperatures), and shown that a single timescale governs all aspects of these dynamics. This framework is applicable to other microorganisms in balanced growth conditions, including those that divide asymmetrically. Moreover, for {\em C. crescentus}, a model organism, we provide a route for determining  the swarmer-to-stalked cell timescale, an important timescale in its life cycle, which was experimentally inaccessible for cells in balanced growth.

The results in [6, 19] indicated that a single condition-specific timescale, which could be be characterized by exponential growth timescale of {\em individual} cell sizes, governed the statistics of growth and division at the single-cell level. Taken together with the scaling results in Figs.~\ref{fig-age-dist} and \ref{fig-timescales}, they reveal the emergence of a cellular unit of time, which can be calibrated for each growth condition of interest by performing a simple measurement of the {\em population} growth rate. Thus at each growth condition of interest, a simple measurement of the {\em population} growth rate reveals the {\em distributions} of cell ages and cell division times!

In this work we have identified the time-dependent cell age distribution, $G(t, \t)$, as an important ``order parameter'' for describing the far-from-equilibrium state of cells under time varying growth conditions. The analytical framework introduced here could provide a starting point for examining stochastic transient responses to changes in external conditions in different contexts: for instance, following a temperature change, a nutritional shift or a chemical perturbation to molecular regulators of cell cycle progression. Evidently, the {\em functional} change of the age-distribution characterizes the transient dynamics following the perturbation. Our framework also provides a route to characterizing essentially non-Markovian time-dependent cellular phenomena, such as the aging dynamics of individual cells under time varying growth conditions.

\section{Acknowledgments}
We thank Steve Berry, Rudro Biswas, Jon Henry, Norbert Scherer, Michael Vennettelli, and Tom Witten for useful discussions, and Kunaal Joshi for a careful reading of the manuscript. We thank Purdue University Startup Funds, Purdue Research Foundation, and the National Science Foundation (NSF REU grant PHY-1460899, NSF PHY-1305542, and NSF DBI-1300426) for financial support. SI-B thanks the Santa Fe Institute, where part of the work was completed, for hospitality.

\section{Author Contributions}
SI-B conceived of and designed research, developed the theoretical model, performed analytical calculations, and observed scaling behaviors reported; FJ extended analytical results to general time-dependent scenarios under SI-B's guidance; CSW performed image and data analyses and tested models under SI-B's guidance; HG and JR performed simulations under the guidance of ARD and SI-B;  ED and KL performed bulk-culture measurements under the guidance of SI-B and AF; SI-B and SC provided reagents and biological supplies; CSW, JR and ARD and SI-B wrote the paper.


\appendix
\section{Analytical Methods}
\subsection*{Case 1: Symmetric division}
\subsubsection*{Definitions.}
\begin{itemize}
\item
$t$ denotes the observation time.
\item
$\t$ denotes the age of the cell measured from the time since it last divided.
\item
$n(t, \t) d\t \equiv$  number of cells at time $t$ with ages between $\t$ and $\t + d\t$.
\item
${N}(t) \equiv \int_{0}^{\infty} d \t \, n(t, \t) \equiv$ total number of cells present at time $t$.  
\item
$G(t, \t) \equiv {n(t, \t)}/{{N}(t)}$. Thus, $G(t, \t) d\t \equiv$ the fraction of cells at time $t$ with ages between $\t$ and $\t + d\t$.
\item
$\a(\t)$, the division propensity, is probability that a cell of age between $\t$ and $\t+ d\t$ will divide in the interval $d \t$. Note that by construction $\a(\t)$ is independent of $t$.
\item
$P(\t) d \t \equiv$ the probability that a cell does not divide up until $\t$ and then divides between $\t$ and $\t+ d\t$. Therefore, $P(\t) d\t =$ [the probability that a cell does not divide until $\t$]$\times \a(\t) d \t$ = $\le[ 1- \int_{0}^{\t} d \t' P(\t')\ri] \a(\t) d \t$. Thus, $\a(\t) =  P(\t)/\le[1- \int_{0}^{\t} d\t' P(\t')\ri]$.
\end{itemize}
\subsubsection*{Time evolution equations for the age distribution.}
Aging of cells and cell-division events result in temporal evolution of $n(t,\t)$:
\begin{align}
\label{eq-eom-n1}
&n(t+dt, \t)d \t = n(t, \t -dt)d\t - \a(\t)dt \,n(t, \t-dt)d\t. \nn \\
&\implies n(t+dt, \t) - n(t, \t) \nn \\
&= - n(t, \t)  + n(t, \t -dt) - \a(\t)dt \,n(t, \t-dt). \nn \\
&\implies \pd_{t} n(t, \t) = -\pd_{\t} n(t, \t) -  \a(\t) n(t, \t).
\end{align}
The products of cell division, $\n$ per cell, appear as new `just-born' $\t=0$ members of the cell population. Accounting for this contribution to the age group $0<\t<dt$, between times $t$ and $t+dt$, one has
\begin{align}\label{eq-eom-n2}
n(t, 0) = \n \le[ \int_{0}^{\infty} d\t n(t, \t) \a(\t) \ri] \equiv \n \rr(t)N(t).
\end{align}
$\rr(t)$, as defined above, is the cell-averaged rate at which newborn cells result across the entire cell population, while $\a(\t)$ is the propensity of birthing new cells at time $\t$.

Now, for $\n>1$,  cell numbers will increase exponentially and thus will not reach a `steady state'. However, we expect the age distribution, $G(t,\t) = n(t,\t)/N(t)$, to have a steady state in all realistic cases. Therefore the goal is to find its time-evolution equation. In order to achieve this, we first find the total population growth rate, using Eqns.~\eqref{eq-eom-n1} and \eqref{eq-eom-n2}:
\begin{align}\label{eq-ntilde-1}
\pd_{t} N(t) &= \int_{0}^{\infty}\le(\pd_{t} n(t,\t)\ri) d\t  \nn \\
&= n(t,0) -n(t,\infty) - \int_{0}^{\infty} d\t n(t, \t) \a(\t) \nn \\
&= \le(\n - 1 \ri)\rr(t)N(t).
\end{align}
The time evolution equation for $G(t,\t)$ is obtained by differentiating both sides of the identity $n(t,\t) = G(t,\t)N(t)$ with respect to $t$, and then using Eqns.~\eqref{eq-eom-n1}, \eqref{eq-eom-n2} and \eqref{eq-ntilde-1}:
\begin{align}
&\pd_{t}n(t,\t) = N(t) \pd_{t}G(t,\t)+ G(t,\t)\pd_{t}N(t). \nn\\
&\imply \; -\pd_{\t} n(t, \t) -  \a(\t) n(t, \t) \nn \\
&= N(t) \pd_{t}G(t,\t)+ G(t,\t)\le(\n - 1 \ri)\rr(t)N(t).
\end{align}
Dividing throughout by $N(t)$, we obtain the time evolution equation of the age distribution:
\begin{align}\label{eq-G-1}
\pd_{t} G(t, \t) + \pd_{\t} G(t, \t) 
+ \a(\t)G(t, \t) 
\nn \\ + \le(\n - 1 \ri) \rho(t) G(t, \t) = 0.
\end{align}
We can account for newborn daughter cells by dividing Eq.~\eqref{eq-eom-n2} by $N(t)$:
\begin{align}\label{eq-G-2}
G(t, 0) = \n \le[ \int_{0}^{\infty} d\t\, G(t, \t) \a(\t) \ri] \equiv \n \rr(t).
\end{align}

\subsubsection*{Steady state solution.}

In steady state the age distribution, $G(t,\t)$, and consequently the birth-rate, $\rr(t)$, will both be time-independent. These steady-state quantities, denoted by the superscript `*', satisfy $t$-independent versions of Eqs.~\eqref{eq-G-1} and \eqref{eq-G-2}:
\begin{subequations}\label{eq-G-3}
\begin{align}
\pd_{\t} G^{*}(\t) + \le[\a(\t) + \le(\n - 1 \ri) \rho^{*}\ri]G^{*}(\t) &= 0,\\
G^{*}(0) = \n \le[ \int_{0}^{\infty} d\t\, G^{*}(\t) \a(\t) \ri] &\equiv \n \rr^{*}.
\end{align}
\end{subequations}
From Eq.~\eqref{eq-ntilde-1} we have
\begin{align}\label{eq-expgrowth}
\pd_{t} N(t) &= \le(\n - 1 \ri)\rr^{*}N(t) \equiv k N(t),\nn\\
\imply\; N(t) &= N{(0)}e^{k t}.
\end{align}
Thus, as expected, the total cell population grows exponentially with the rate
\begin{align}\label{eq-kappa}
k = \le(\n - 1 \ri)\rr^{*}.
\end{align}
Defining $A(\t) \equiv \int_{0}^{\t} d\t' \a(\t')$ and using the above relation between $k$ and $\rr^{*}$, Eq.~\eqref{eq-G-3} becomes
\begin{align}
\frac{d} {d \t}  \le[ G^{*}(\t) \, e^{k \t + A(\t)} \ri] = 0.
\end{align}
Integrating both sides from $\t'=0$ to $\t' = \t$, and using Eq.~\eqref{eq-G-2},
\begin{align}
G^{*}(\t) \, e^{k\t + A(\t)} =  G^{*}(0) = \n \rho^{*} = \frac{k \n}{\n - 1}.
\end{align}
Thus, the steady state age distribution is 
\begin{align}
\label{eq-G-soln-1}
{G^{*}(\t) = \frac{k \n }{\n-1}\, e^{-k\t}e^{-A(\t)}}.
\end{align}
When $\n>1$, the only unknown number in this expression, $k$, may be found from the normalization of the probability density, $G^{*}(\t)$:
\begin{align}
1 = \int_{0}^{\infty} d\t \,G^{*}(\t) = \frac{k \n }{\n-1}\int_{0}^{\infty} d\t\; e^{-k\t}e^{-A(\t)}.
\end{align}
Using the relation $P(\t) = \a(\t)e^{-A(\t)}$, which follows from the definitions of $\a(\t)$ and $A(\t)$,
the normalization equation for $G^{*}$ becomes
\begin{align}\label{eq-expgrowth-exponent}
\int_{0}^{\infty} d\t P(\t) e^{-k\t} \equiv {\langle e^{-k\t} \rangle}_{P} = \frac{1}{\n}. 
\end{align}
When $\n=1$, system is closed, since the total number of cells is conserved. For this case, since $N(t)$ is a constant, $\rho$ is a constant, and a simplification of the previous general derivation is obtained. 
Therefore Eq.~\eqref{eq-G-3} can be directly integrated to compute the normalized steady-state age distribution for $\n = 1$:
\begin{align}
\label{eq-start}
G^{{*}}(\t) = 1 - \frac{1}{\m_{\t}}\int_{0}^{\t} d\t' P(\t') \mbox{  } \equiv \frac{1}{\m_{\t}}\int_{\t}^{\infty}d\t' P(\t'),
\end{align}
where the mean value, $\m_{\t}$, is evaluated w.r.t. the division time distribution, $P(\t)$.

Using this expression, the mean age, $\m_{a}$, for the case $\n=1$ can be evaluated:
\begin{align}
\label{eq-shady}
\m_{a} = \int_{0}^{\infty} d\t \, G^{{*}}(\t) =  \frac{1}{\m_{\t}}\int_{0}^{\infty} d\t \, \int_{\t}^{\infty}d\t' P(\t').
\end{align}
\begin{figure}[h]
\begin{center}
\includegraphics[width= 0.5\columnwidth]{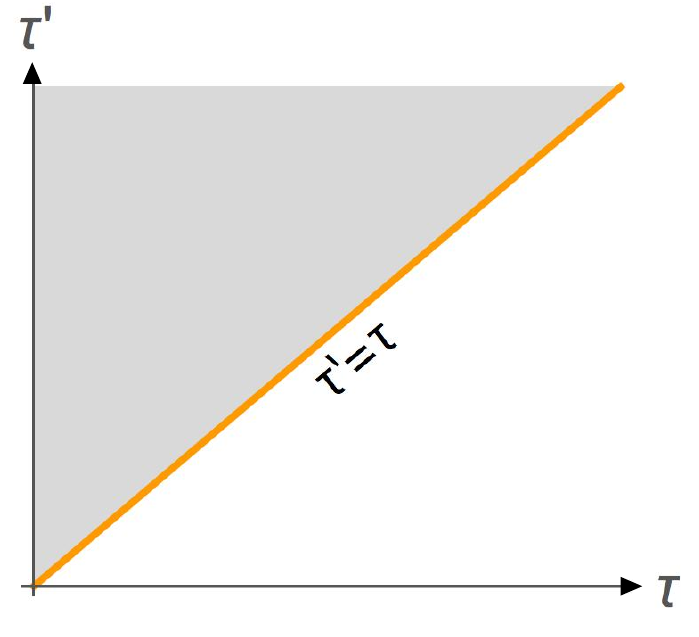}
\caption{{\bf Region of integration for Eq.~\eqref{eq-shady}}. The shaded area corresponds to the region of integration in the $\t-\t'$ plane, for the expression on the right hand side of Eq.~\eqref{eq-shady}.}
\label{fig-doubleintegral}
\end{center}
\end{figure}
The double integral is over the shaded region in the $\t-\t'$ plane as shown in Fig.~\ref{fig-doubleintegral}, and the order of integrals may be interchanged as follows:
\begin{align}
\label{eq-stop}
\m_{a} &=  \frac{1}{\m_{\t}}\int_{0}^{\infty} d\t' P(\t') \, \int_{0}^{\t'}d\t \, \t\nn\\
&= \frac{1}{\m_{\t}} \int_{0}^{\infty} d\t' P(\t') \frac{(\t')^{2}}{2}\nn\\
&= \frac{\s_{\t}^{2} + \m_{\t}^{2}}{2\m_{\t}} = \frac{\m_{\t}}{2}\le(1 + \eta_{\t}^{2} \ri).
\end{align}
In this equation, $\s_{\t}^{2}$ denotes the variance of $\t$ while $\eta_{\t} = \s_{\t}/\m_{\t}$ is the corresponding coefficient of variance. This is the derivation of {Eq.~\eqref{eq-means}} in the main text.

\subsection*{Case 2:  Asymmetric division}
\subsubsection*{Time evolution equations for the age distributions \\of normal and quiescent cells.}
The cell population now has two distinct cell types: normal reproducing cells, and quiescent cells which transition to normal cells before they can divide. The normal cells divide with propensity $\a(\t)$ at age $\t$, creating $\n$ normal and $\n_{q}$ quiescent cells. Their division time distribution is $P(\t) = \a(\t)e^{-A(\t)}$, where $A(\t)=\int_{0}^{\t}\a(\t')d\t'$. Thus, these terms are defined just as we did for the symmetric case. The quiescent cells are similarly defined, transitioning to normal cells with propensity $\a_{q}(T_{q})$ and with a corresponding waiting time distribution $P_{q}(T_{q}) = \a_{q}(T_{q})e^{-A_{q}(T_{q})}$, where $A_{q}(T_{q})=\int_{0}^{T_{q}}\a_{q}(T_{q}')\,d T_{q}'$. Using the subscript $q$ to denote the quantities defined for the quiescent cells, the time evolution equations for the number density of cells with age $\t$ are:
\begin{subequations}\label{eq-asym-eom-n}
\begin{align}
\pd_{t} n(t, \t) &= -\pd_{\t} n(t, \t) -  \a(\t) n(t, \t),\\
\pd_{t} n_{q}(t, T_{q}) &= -\pd_{T_{q}} n_{q}(t, T_{q}) -  \a_{q}(T_{q}) n_{q}(T_{q}).
\end{align}
\end{subequations}
Newborn  cells result from  cell division of normal cells, as well as from the conversion of quiescent to normal cells:
\begin{subequations}\label{eq-asym-n0}
\begin{align}
n(t,0) &= \n N(t) \rr(t) + N_{q}(t) \rr_{q}(t),\\
n_{q}(t,0) &= \n_{q} N(t) \rr(t).
\end{align}
\end{subequations}
Analogous to the symmetric case, the per-cell division and conversion rates, $\rr$ and $\rr_{q}$, are defined as follows:
\begin{subequations}\label{eq-asym-rho}
\begin{align}
\rr(t)&=\int_{0}^{\infty} d\t \frac{n(t, \t)}{N(t)} \a(\t) = \int_{0}^{\infty} d\t \, G(t, \t) \a(\t),\\
\rr_{q}(t)&=\int_{0}^{\infty} d T_{q} \frac{n_{q}(t, T_{q})}{N_{q}(t)} \a_{q}(T_{q}) \nn \\ &= \int_{0}^{\infty} dT_{q} \,G_{q}(t, \t_{q}) \a_{q}(\t_{q}).
\end{align}
\end{subequations}
Combining these equations yields the time evolution equations for total population numbers $N(t)$ and $N_{q}(t)$:
\begin{subequations}\label{eq-asym-tot-n}
\begin{align}
\pd_{t}N &= (\n -1)N \rr + N_{q}\rr_{q},\\
\pd_{t}N_{q} &= \n_{q} N \rr - N_{q}\rr_{q}.
\end{align}
\end{subequations}

\subsubsection*{Steady state solution.}

After a sufficiently long time, we expect the age distributions $G$ and $G_{q}$, and thus the cell-averaged division and conversion rates, respectively $\rr$ and $\rr_{q}$, to stabilize and become time independent. Below we consider only this long time steady state limit of these quantities, and as such we discard the previous use of the superscript '$*$' in this context. For later use, we define the following (constant) ratio in steady state:
\begin{align}\label{eq-asym-gamma}
\g = \frac{\rr_{q}}{\rr}.
\end{align}
In steady state we expect both cell population numbers to grow exponentially with the \emph{same} rate $k$:
\begin{align}\label{eq-asym-kappa}
\frac{\pd_{t}N(t)}{N(t)} = \frac{\pd_{t}N_{q}(t)}{N_{q}(t)} = k, \quad \imply \quad N(t), N_{q}(t) \propto e^{k t}.
\end{align}
From this, it is clear that the following ratio must become time-independent:
\begin{align}\label{eq-asym-phi}
\f = \frac{N_{q}(t)}{N(t)}.
\end{align}
Eqns.~\eqref{eq-asym-tot-n}, \eqref{eq-asym-gamma} and \eqref{eq-asym-phi} can be combined to yield the following steady state equations:
\begin{subequations}\label{eq-asym-tot-n-ss}
\begin{align}
\frac{\pd_{t}N(t)}{N(t)} &= \rr\le(\n + \f \g - 1\ri),\\
\frac{\pd_{t}N_{q}(t)}{N_{q}(t)} &= \rr \le(\frac{\n_{q} - \f\g}{\f}\ri).
\end{align}
\end{subequations}
Comparing these with Eq.~\eqref{eq-asym-kappa}, we find
\begin{align}\label{eq-asym-kappabyrho}
\frac{k}{\rr} = \n + \f \g - 1 = \frac{\n_{q} - \f\g}{\f}.
\end{align}
Combining Eqs.~\eqref{eq-asym-eom-n}, \eqref{eq-asym-n0} and \eqref{eq-asym-kappa}, we can write down the equations satisfied by the age distributions $G$ and $G_{q}$ in steady state:
\begin{subequations}\label{eq-asym-eom-G}
\begin{align}
\pd_{\t}G(\t) + \le[k + \a(\t)\ri]G(\t) &= 0,\\
\pd_{\t}G_{q}(\t) + \le[k + \a_{q}(\t)\ri]G_{q}(\t) &= 0.
\end{align}
\end{subequations}
The initial conditions are, respectively, $G(0) = \rr (\n + \f\g)$ and $G_{q}(0) = {\rr \n_{q}}/{\f}$. These equations are then solved to obtain the expressions:
\begin{subequations}\label{eq-asym-G-sol}
\begin{align}
G(\t) &= \rr (\n + \f\g) e^{-k\t}e^{-A(\t)}, \;  \; \nn \\ A(\t) &= \int_{0}^{\t} \a(\t^{'})\,d\t^{'},\\
G_{q}(T_{q}) &= \frac{\rr \n_{q}}{\f} e^{-k T_{q}}e^{-A_{q}(T_{q})},  \;  \; \nn \\ \quad A_{q}(T_{q}) &= \int_{0}^{T_{q}} \a_{q}(T_{q}^{'})\,dT_{q}^{'}.
\end{align}
\end{subequations}
Normalizing the two distributions $G$ and $G_{q}$, analogous to the symmetric case above, we find:
\begin{subequations}\label{eq-asym-G-normalize}
\begin{align}
{\langle e^{-k \t} \rangle}_{P} &= 1 - \frac{k}{\rr(\n + \f\g)} = \frac{1}{\n + \f\g},\\
{\langle e^{-k T_{q}} \rangle}_{P_{q}} &= 1 - \frac{k \f}{\rr \n_{q}} = \frac{\g\f}{\n_{q}}.
\end{align}
\end{subequations}
To derive these equations we have used Eq.~\eqref{eq-asym-kappabyrho}. Eliminating $\g\f$ from these equations, we find the equation (analogous to Eq.~\eqref{eq-expgrowth-exponent} for the symmetric case) which determines the growth rate, $k$:
\begin{align}\label{eq-asym-kappa-sol}
{{\langle e^{-k \t} \rangle}_{P}  = \frac{1}{\n + \n_{q} {\langle e^{-k T_{q}} \rangle}_{P_{q}}}}.
\end{align}
Using this in Eq.~\eqref{eq-asym-G-normalize}, one can determine the value of $\f\g$. These, combined with Eq.~\eqref{eq-asym-kappabyrho}, yield the individual values of $\rr$, $\f$ and $\g$, thus solving the full steady state problem. \\

\subsubsection*{General time-dependent solution.}
We have obtained the complete time-dependent solution to the asymmetric cell division process specified by Eqs.\ \eqref{eq-asym-eom-n}, \eqref{eq-asym-n0} and \eqref{eq-asym-rho}, in terms of the initial population distributions,
\begin{align}
n^{(0)}(\t) = n(0,\t), \; n^{(0)}_{q}(T_{q}) = n_{q}(0, T_{q}).
\end{align}
To derive the solution we proceed as follows. Using the method of characteristics, we obtain from Eq.~\eqref{eq-asym-eom-n}: 
\begin{subequations}\label{eq-time-delay}
\begin{align}
	&n(t, \tau) = \begin{cases}
		n(t-\tau,0) f(\tau)	&	\tau < t\\
		n^{(0)}(\tau-t) \frac{f(\tau)}{f(\tau-t)}		&	\tau \geq t
	\end{cases},\\
	&n_q(t, T_{q}) = \begin{cases}
		n_q(t-T_{q},0) f_q(T_{q})	&	T_{q} < t\\
		n^{(0)}_q(T_{q}-t) \frac{f_q(T_{q})}{f_q(T_{q}-t)}		&	T_{q} \geq t
	\end{cases},
\end{align}
\end{subequations}
where, $f(\tau)$ and $f_q(T_{q})$ are defined as:
\begin{subequations}
\begin{align}
	&f(\tau) = e^{-\int_0^\tau\alpha(\tau')d\tau'},\\
	&f_q(T_{q}) = e^{-\int_0^{T_{q}}\alpha_q(T'_{q})dT'_{q}}.
\end{align}
\end{subequations}
$f(\tau)$ and $f_q(T_{q})$ respectively represent the probabilities that a normal (quiescent) cell will not divide (differentiate) by age $\tau$ ($T_{q}$). These equations imply the following.
\begin{itemize}
\item
When $\tau<t$ ($T_{q}<t$), the number of normal (quiescent) cells of age $\tau$ ($T_{q}$) at time $t$ is given by the number of cells of age $0$ at time $t-\tau$ ($t - T_{q}$), multiplied by the probability that they have not divided (differentiated) by age $\tau$ ($T_{q}$);
\item
When $\tau\geq t$ ($T_{q}>t$), the number of normal (quiescent) cells of age $\tau$ ($T_{q}$) at time $t$ is given by the initial number of cells of age $\tau-t$ ($T_{q}-t$), multiplied by the probability that they have not divided (differentiated) between the ages of $\tau - t$ ($T_{q}-t$) and $\tau$ ($T_{q}$).
\end{itemize}
These expressions still involve populations of cells with age equal to $0$ on the right hand side.  We can eliminate them using Eqs.~\eqref{eq-asym-n0} and \eqref{eq-asym-rho}. Defining these populations as
\begin{align}\label{eq-NBdefs}
n^{NB}(t) \equiv n(t,0), \; n^{NB}_{q}(t) \equiv n_{q}(t, 0),
\end{align}
and combining Eqs.~\eqref{eq-time-delay}, \eqref{eq-asym-n0} and \eqref{eq-asym-rho}, we obtain:
\begin{subequations}\label{eq-newborns}
\begin{align}
n^{NB}(t) &= \n (n^{NB}*P)(t)+\n n_{1}(t) + (n^{NB}_{q}*P_{q})(t) + n_{2}(t),\\
n^{NB}_{q}(t) &= \n_{q} (n^{NB}*P)(t) + \n_{q} (n^{NB}*P)(t) n_{1}(t).
\end{align}
\end{subequations}
In these expressions we have used the notation `$g*f$' to denote a convolution between functions $g(t)$ and $f(t)$:
\begin{align}
(g*f)(t)  = \int_{0}^{t}g(t-t')f(t') dt'.
\end{align}
The functions $n_{1,2}$ are defined in terms of the initial population as follows:
\begin{subequations}
\begin{align}
n_{1}(t)  = \int_{t}^{\infty} n^{(0)}(\t - t)\frac{f(\t)}{f(\t - t)}\a(\t) d\t,\\
n_{2}(t)  = \int_{t}^{\infty} n^{(0)}_{q}(T_{q} - t)\frac{f_{q}(T_{q})}{f_{q}(T_{q} - t)}\a(\t) dT_{q}.
\end{align}
\end{subequations}

Using the Laplace transform, Eq.~\eqref{eq-newborns} can be inverted to obtain the time-dependent populations with age $0$ (Eq.~\eqref{eq-NBdefs}), in terms of quantities that are explicit functions of the initial cell populations:
\begin{subequations}
\label{eq-n0}
\begin{align}
n(t,0) &=  \mathcal L^{-1}\left\{\frac{\tilde n_2(s) +\tilde n_1(s)(\nu +\nu_q P_q(s))}{1-\tilde P(s) (\nu +\nu_q \tilde P_q(s))}\right\}(t),\\
n_q(t,0) &= \mathcal L^{-1}\left\{\nu_q\frac{\tilde n_2(s)\tilde P(s)+\tilde n_1(s)}{1-\tilde P(s) (\nu +\nu_q \tilde P_q(s))} \right\}(t).
\end{align}
\end{subequations}
In these expressions $\mathcal L^{-1}$ denotes the Inverse Laplace Transform. Finally, these expressions can be substituted in Eq.~\eqref{eq-time-delay} to obtain the complete time-dependent solution for the population age distributions for any initial population distribution.

Integrating this solution over all ages, we find expressions for the total population numbers of normal and quiescent cells at any time $t$:
\begin{subequations}
\label{eq-totnum}
\begin{align}
N(t) &= \mathcal L^{-1} \left\{ \tilde f(s) \frac{\tilde n_2(s) +\tilde n_1(s)(\nu +\nu_q P_q(s))}{1-\tilde P(s) (\nu +\nu_q \tilde P_q(s))}\right\}(t) \nn\\
& \quad + \int_t^\infty n^{(0)}(\tau-t) \frac{f(\tau)}{f(\tau-t)} d\tau,\\
N_{q}(t) &=  \mathcal L^{-1} \left\{ \tilde f_q(s) \nu_q \frac{\tilde n_2(s)\tilde P(s)+\tilde n_1(s)}{1-\tilde P(s) (\nu +\nu_q \tilde P_q(s))}\right\}(t) \nn\\
& \quad + \int_t^\infty n^{(0)}_q(T_{q}-t) \frac{f_q(T_{q})}{f_q(T_{q}-t)} dT_{q}.
\end{align}
\end{subequations}

The time-dependent age distributions for normal and quiescent cells, respectively $G(t, \t)$ and $G_{q}(t, T_{q})$, are equal to the ratio of their corresponding number densities, $n(t, \t)$ and $n_{q}(t, T_{q})$, with the corresponding total numbers, $N(t)$ and $N_{q}(t)$. Thus, they can be computed using the expressions in Eqs.~\eqref{eq-time-delay}, \eqref{eq-n0} and \eqref{eq-totnum}.


%

\clearpage


\begin{center}
{\bf\Large Supplemental Information}
\end{center}
\setcounter{secnumdepth}{3}  
\setcounter{section}{0}
\setcounter{equation}{0}
\setcounter{figure}{0}
\renewcommand{\theequation}{S-\arabic{equation}}
\renewcommand{\thefigure}{S\arabic{figure}}
\renewcommand\figurename{Supplementary Figure}
\renewcommand\tablename{Supplementary Table}
\newcommand\Scite[1]{[S\citealp{#1}]}
\makeatletter \renewcommand\@biblabel[1]{[S#1]} \makeatother

\begin{figure}[h]
\begin{center}
\resizebox{14cm}{!}{\includegraphics{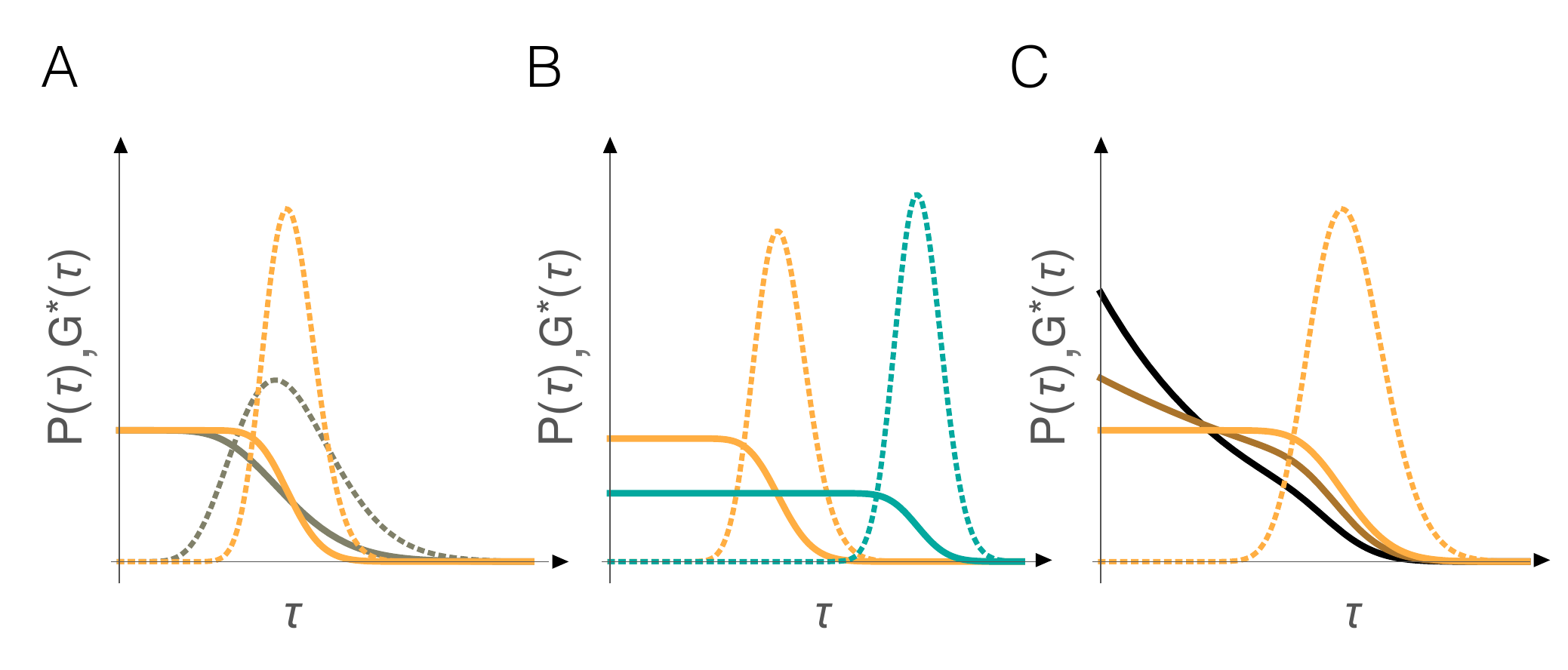}}
\caption{{\bf Relation between the division time distribution, $P(\t)$, and the steady-state age distribution, $G^{*}(\t)$, for a given progeny number, $\n$.} See main text, Eqs. (1), (2) and (3) and accompanying text, for explanation of symbols and discussion of results. In (A) we have shown cell-age distributions (bold orange and gray curves) for $\n=1$ corresponding to two $P(\t)$ distributions (dotted orange and gray); the $P(\t)$ distributions have the same mean but different COVs.  The figure illustrates that the point of inflection of the cell-age distribution determines the mean division time and the slope at the point of inflection determines the width of the division time distribution. In (B) we have shown cell-age distributions (bold orange and cyan) for $\n=1$, corresponding to two $P(\t)$ distributions (dotted orange and cyan) which have the same COV but different means. Evidently, the point of inflection of $G^{*}(\t)$ predicts the mean  division time. In (C) we show age distributions for $\n = 1$ (bold orange), $\n= 2$ (brown) and $\n = 5$ (black) for the same division time distribution (dotted orange).}
\label{fig-supp1}
\end{center}
\end{figure}

\href{http://iyerbiswas.com/BridgingScales/SupplementaryVideo1.mov}{\textcolor{MidnightBlue}{\bf Supplementary Video 1}}
\\

\href{http://iyerbiswas.com/BridgingScales/SupplementaryVideo2.mov}{\textcolor{MidnightBlue}{\bf Supplementary Video 2}}
\\

\end{document}